\documentclass[letterpaper,twocolumn,10pt]{article}
\usepackage{usenix,epsfig,endnotes}
\usepackage{amsmath}
\usepackage{amsthm}
\usepackage{amsfonts}
\usepackage{amssymb}
\usepackage{graphicx}
\usepackage{url}
\usepackage{color}    
\usepackage{cite}
\usepackage[ruled]{algorithm2e}
\usepackage{xspace}
\usepackage{hyperref}
\usepackage{msc}
\usepackage{listings}
\usepackage{paralist}
\usepackage{tikz}
\usepackage{moreverb}

\usetikzlibrary{decorations.pathreplacing}

\newcommand{\namelong}{Persistent and Accountable Domain Validation\xspace}
\newcommand{\name}{PADVA\xspace}

\begin{document}
\title{Blockchain-based TLS Notary Service}
\author{{\rm Pawel Szalachowski}\\
    SUTD\\pawel@sutd.edu.sg
}
\date{}


\maketitle

\begin{abstract}
    The Transport Layer Security (TLS) protocol is a de facto standard of secure
    client-server communication on the Internet.  Its security can be diminished
    by a variety of attacks that leverage on weaknesses in its design and
    implementations.  An example of a major weakness is the public-key
    infrastructure (PKI) that TLS deploys, which is a \textit{weakest-link}
    system and introduces hundreds of \textit{links} (i.e., trusted entities).
    Consequently, an adversary compromising a single trusted entity can
    impersonate any website.
    
    Notary systems, based on multi-path probing, were early and promising
    proposals to detect and prevent such attacks.  Unfortunately, despite their
    benefits, they are not widely deployed, mainly due to their long-standing
    unresolved problems.  In this paper, we present \namelong (\name), which is a
    next-generation TLS notary service.  \name combines the advantages of
    previous proposals, enhancing them, introducing novel mechanisms, and
    leveraging a blockchain platform which provides new features.  \name keeps
    notaries auditable and accountable, introduces service-level agreements and
    mechanisms to enforce them, relaxes availability requirements for notaries,
    and works with the legacy TLS ecosystem.  We implemented and evaluated
    \name, and our experiments indicate its efficiency and deployability.
\end{abstract}

\section{Introduction}
\label{sec:intro}
The security of TLS connections strongly depends on the authenticity of public
keys.  In the TLS PKI~\cite{rfc5280}, a public key is authenticated by a trusted
certification authority (CA), whose task is to verify a binding between an
identity and a public key, and subsequently, issue a certificate asserting it.
Unfortunately, this process does not provide a high-security level for various
reasons. The identity verification is usually conducted over the Internet basing
on the \textit{trust on first use} model~\cite{bhargavan2017formal}, moreover,
it is a one-time operation (i.e., per certificate issuance). Therefore, an
adversary able to impersonate a domain, even for just a moment, can obtain a
valid certificate for this domain.  Besides that, the TLS PKI is a weakest-link
system, and compromising a single CA (out of hundreds of trusted
CAs~\cite{perl2014you}) can result in a successful impersonation attack, as
observed in the past~\cite{MozillaDigiNotarRemoval,comodo}.

One of the first approaches to mitigate such attacks were notary
systems~\cite{wendlandt2008perspectives,marlinspikeconvergence}.  The main idea
behind them is to introduce the new trusted party, known as a notary. The notary
provides a TLS client with its view of the contacted server's public key. Hence,
the client gets better guarantees that the key is legitimate.  Notary systems
are based on multi-path probing and assume that attacks are usually short-lived
or/and scoped to a network topology fragment.  Although they inspired the
research community, they did not receive enough traction to be widely deployed.
Notary systems are often critiqued as they introduce privacy issues (i.e., TLS
clients reveal servers they want to contact to), increase latency of TLS
connections (a response from a notary has to be delivered), and are required to
provide high availability (otherwise client queries will timeout or introduce
significant latency)~\cite{not_convergence,bates2014forced,merzdovnik2016whom}.
Furthermore, notaries are trusted, not transparent, and difficult to audit.

In this paper, we present \namelong (\name), a next-generation TLS notary
service that solves multiple problems of the previously proposed notary systems.
In \name, notaries persistently validate public keys of domains in an auditable
and accountable manner.  By leveraging properties of a blockchain platform,
\name achieves transparency, provides a framework for service-level agreement
(SLA) enforcement, and relaxes availability requirements.  \name is compatible
with legacy TLS infrastructure, and can be deployed today, even with TLS servers
with desynchronized clocks.  Our implementation and evaluation indicate
efficiency and deployability of our system.  With \name, the key validation becomes
a service (instead of a one-time operation) with implementable SLAs, and
with auditable and accountable notaries.

\section{Background and Preliminaries}
\label{sec:pre}
In this section, we introduce basic information on the TLS protocol, we define
the problem, and introduce system and adversary models.

\subsection{Transport Layer Security}
\label{sec:pre:tls}
The TLS protocol is designed to provide secure communication in the
client-server model. It is widely deployed and its most prominent use is to
secure the HTTP protocol (HTTPS).  In such a setting, only a (web) server is
authenticated and the authentication is based on X.509
certificates~\cite{rfc5280}. The current and the recommended version of the
protocol is TLS 1.2~\cite{rfc5246}, however, older versions, like 1.0 and 1.1,
are still widely supported.\footnote{\url{https://www.ssllabs.com/ssl-pulse/}}

\subsubsection{Handshake}
\label{sec:pre:tls:handshake}
The first phase of the TLS connection establishment is the TLS handshake protocol.
Its goal is to securely negotiate a shared symmetric key between communicating
parties.  There are many variants of the TLS handshake and we present the
version based on the ephemeral Diffie-Hellman (DH)
protocol~\cite{blake1998authenticated}. This version is recommended due to its
security benefits (i.e., it provides forward secrecy), and is used by default by
modern TLS clients~\cite{whytls13} and
servers.\footnote{\url{https://www.ssllabs.com/ssltest/clients.html}}

\begin{figure}[h!]
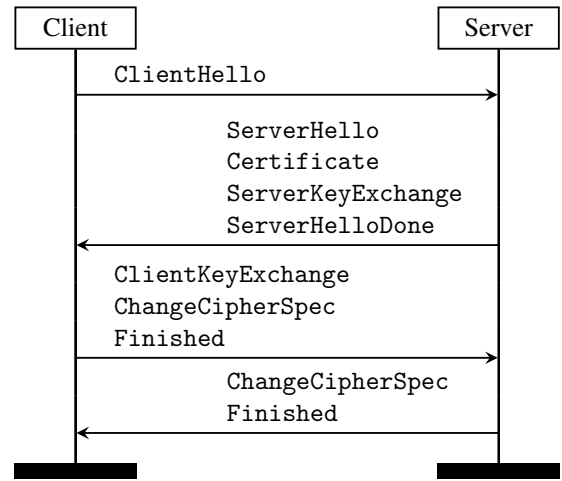

    \vspace{-1cm}
    \centering
    \drawframe{no}
    \setmsckeyword{}
    \setlength{\envinstdist}{1.0cm}
    \setlength{\instdist}{4.0cm}
    \begin{msc}{}
        \small
        \declinst{cli}{}{Client}
        \declinst{ser}{}{Server}

        \mess{
            \parbox{4.6cm}{
                \texttt{ClientHello}
            }
        }{cli}{ser}
        \nextlevel
        \nextlevel
        \nextlevel
        \nextlevel
        \mess{
            \parbox{1.6cm}{
                \texttt{ServerHello\\Certificate\\ServerKeyExchange\\ServerHelloDone}
            }
        }{ser}{cli}
        \nextlevel
        \nextlevel
        \nextlevel
        \mess{
            \parbox{4.6cm}{
                \texttt{ClientKeyExchange\\ChangeCipherSpec\\Finished}
            }
        }{cli}{ser}
        \nextlevel
        \nextlevel
        \mess{
            \parbox{1.6cm}{
                \texttt{ChangeCipherSpec\\Finished}
            }
        }{ser}{cli}

    \end{msc}
    \vspace{-0.25cm}
    \caption{The TLS handshake with an ephemeral DH key exchange.}
    \label{fig:tls_handshake}
\end{figure}

The handshake is presented in \autoref{fig:tls_handshake}.  It starts with
exchanging the \texttt{ClientHello} and \texttt{ServerHello} messages, where the
client and server indicate their supported cryptographic primitives, their
random nonces, and other connection parameters.  Next, the server sends its
certificate and the \texttt{ServerKeyExchange} message.  This message contains
DH parameters required to conduct an ephemeral DH key exchange and a signature
that protects these parameters. This signature is computed using the private key
corresponding to the public key of the server's certificate.  After the
ServerHello phase is finished, the client verifies the signature and sends its
DH contribution. Next, the client and server compute the shared secret and will
then signal that the following communication will be encrypted.

\subsubsection{Certificates}
\label{sec:pre:tls:cert}
Certificates in the TLS protocol are used to verify end entities.  In most
cases, like HTTPS, only the server is authenticated (i.e., the client does not
have a certificate), and usually, the server is identified by its domain name.
Certificates are issued by a trusted CA that is obligated to validate the
ownership of a public key prior to the certificate issuance.  In the current TLS
ecosystem, the dominant way of the validation is based on a domain name
ownership.  To get a domain-validated certificate, an entity has to prove to a
CA that it controls the domain name for which a certificate is requested.  CAs
automate this process, and usually the validation is conducted via DNS, HTTP,
or e-mail~\cite{bhargavan2017formal}.

\subsection{Problem Definition}
\label{sec:pre:prob_def}
Notary systems were motivated by the weak authentication of public keys that the
standard TLS PKI model provides.  As described above, in
most cases, CAs rely only on a domain validation conducted over an insecure
channel, moreover, they validate the given domain only once per certificate's
lifetime (i.e., just before the certificate is issued). Consequently, an
adversary with the ability to temporarily launch a man-in-the-middle (MitM) attack
between any CA and a domain, or temporarily control the domain's DNS zone,
e-mail, or a web server, can get a valid certificate for this domain.  Such a
certificate can be then used to impersonate the domain to any client.

Notary systems try to improve the security of the TLS connections, providing their
views of domains' public keys.  Notaries can periodically check domains'
public keys and the \textit{key continuity}~\cite{key-cont} of the observed keys can
be measured, such that notaries can inform clients about historical views to
give them better guarantees about public keys they currently see.  The effectiveness
of notary systems is based on the following assumptions:
\begin{compactitem}
    \item impersonation attacks are usually short-lived, thus key continuity,
        that says for how long a public key is being used by an entity,   can be
        a practical measure for estimating how a given public key is
        \textit{suspicious},
    \item multipath probing can detect various MitM attacks, as usually attacks
        are limited to a fragment of the network topology.
\end{compactitem}
\medskip

Unfortunately, notary systems  never received mainstream deployment, mainly due to
availability, privacy, and security issues.  Below, we list the desired
properties that a successful notary system should hold. This list is motivated
by lessons learned from deployments of notary (and other security)
infrastructures.
\begin{compactdesc}
    \item[Persistence:] in contrast to today CAs, notaries should validate
        public keys persistently.  The validation conducted persistently can
        increase security level, as then an adversary has to have a permanent
        ability to impersonate a targeted domain.  With such validation, it
        is possible to reliably implement security metrics like key
        continuity.  Interestingly, similar benefits can be provided by
        short-lived certificates~\cite{rivest1998can}, however, due to the CA
        ecosystem ossification the attempts of introducing short-lived
        certificates
        failed.\footnote{\url{https://cabforum.org/2015/11/11/ballot-153-short-lived-certificates/}}
    \item[Auditability:] in the previous proposals notaries are trusted.  Such a
        setting is circular as trust problems of the CA ecosystem are addressed
        by another trusted parties (notaries).  Some level of trust seems to be
        unavoidable as clients that contact a notary have to rely on the notary's
        view. However, where possible notaries should be auditable, such that
        they can be kept accountable for their actions.  Ideally, operations of
        notaries are authentic and transparent, such that anyone can audit them.
    \item[Privacy:] notary systems are designed in the interactive client-server
        model, where clients wishing to check notary's view have to contact
        notary server(s).  That design gives a notary infrastructure ability to
        learn which websites are visited by which clients. This is unacceptable
        and a successful notary system should be privacy-preserving.
    \item[Availability:] another consequence of the interactive model is an
        increased latency of TLS connection establishments and strong
        availability requirements, as information from a notary has to be
        reliably sent to clients.  Unfortunately, maintaining high availability
        level for front-end servers is a challenging task, and currently,
        existing security infrastructures (like PKI revocation
        infrastructures~\cite{rfc5280} or certificates logs~\cite{rfc6962}) have
        major problems with achieving
        it~\cite{liu2015emc,ct_uptime,ct_uptime2,ct_old_sth}.
    \item[Backward compatibility:] a notary should be compatible with the legacy
        TLS ecosystem (or at least a large its portion), such that it can
        provide benefits immediately when deployed.  The TLS PKI ecosystem has
        many stakeholders like CAs, browser vendors, and influential websites.
        As we witnessed in the past, it is challenging to deploy a new security
        enhancement without reaching a broad
        consensus~\cite{matsumoto2015deployment}.  Ideally, a notary system
        should be orthogonal to the deployed TLS protocol and the TLS PKI, such
        that it can be deployed almost immediately without requiring flag days,
        servers' upgrades, and protocol changes.
\end{compactdesc}

\subsection{Assumptions}
\label{sec:pre:assumptions}

\subsubsection{System Model}
\label{sec:pre:assumptions:system}

\name introduces the following parties:
\begin{compactdesc}
    \item[Server:] provides a service via the TLS protocol (e.g., HTTPS). We
        assume that services (and servers consequently) are identified by domain
        names.
    \item[Notary:] is an entity that offers a \name service. Notaries are
        obligated to monitor servers by validating their public keys.
    \item[Requester:] is an entity interested in monitoring server's public
        key(s). It can be a server's operator or any other entity.  A requester
        orders the \name service from a notary.
\end{compactdesc}
\medskip

We assume that notaries and requesters have an access to a blockchain platform
with smart contracts enabled.  For simple description, we also assume that the
platform has its cryptocurrency that is used by notaries and requesters, however money
transfers in \name can be realized with a fiat currency if desired (e.g., due to
cryptocurrency volatility).  For instance, our scheme can be combined with
systems like Ethereum\footnote{\url{https://www.ethereum.org/}} or
HyperLedger\footnote{\url{https://hyperledger.org/}}.  We describe \name using
an open blockchain platform, however, \name can be adjusted to a private platform
if needed.  We do not require that clocks of servers are synchronized. However,
we assume that the requester and notary can agree on at least one reference time
source (in particular this can be the monitored server).

\subsubsection{Adversary Model}
\label{sec:pre:assumptions:adversary}
With regards to MitM attacks, we use the adversary model introduced by previous
notary systems~\cite{wendlandt2008perspectives,marlinspikeconvergence}. Namely,
we assume that   impersonation attacks have limited duration, or are limited to
only a fragment of the network topology.  Otherwise, a long-lived attack that
encompasses the entire topology would be difficult to detect by any notary
system (as the detection bases on finding views' inconsistency).

We assume that a notary can misbehave by reporting an invalid view or censoring
requester audit queries.  However, in such a case the misbehavior should be
detected and the notary should be punished. 

We assume that the adversary cannot undermine security properties of the
underlying blockchain platform, and we assume that the adversary cannot
compromise the security of the used cryptographic primitives.

\section{\name Overview}
\label{sec:overview}
The main observation behind the \name's design  is that although benefits of
notary systems are attractive, their deployment is marginal due to deployment,
operational, and design issues.  Hence, first we introduce the core design
decisions that enable to build a more powerful notary system, and then we
outline an overview of \name.

\subsection{Design Choices}
Our first design choice is to position a \textbf{persistent and auditable key
validation} as the main task of notaries.   In \name, notaries constantly
validate public keys of domains and measure key continuity.  That allows
implementing effective protection and detection mechanisms in an ecosystem with
predominant domain-validated certificates.  In such a setting, there are some
similarities between notaries and CAs conducting domain validation except:
\begin{compactitem}
    \item notaries validate public key of domains permanently (in short intervals),
        and
    \item notaries are not authorized to issue certificates.
\end{compactitem}
Besides that, \name aims to provide auditable and accountable key validation.
Notaries in our system are not trusted parties, as they have to get a proof that
given domain used given key at given time. These proofs, called
\textit{validation results}, and other actions of notaries are
auditable what enables requesters to verify them.

\name notaries offer \textbf{key validation as a service}.  In short, we
argue that notaries should serve a demand-driven service, ideally with payment and SLA
frameworks available.  There are many challenges associated with designing and
implementing key validation as a service. However, offering key validation as a
service would incentivize notaries to maintain a robust infrastructure.
On the other hands, implementable SLAs would secure requesters from notaries
violating mutual agreements.

To be deployed, a notary system must be \textbf{compatible with the legacy
ecosystem}.  It means that we should not propose any modifications to the TLS
protocol nor need any server-side changes (including even small changes like
time synchronization).  The system should be deployable within the current
environment.

This choice is motivated by the history of TLS security enhancements deployment.
Many security proposals failed due to server-side changes required or due to low
adoption rate (as to work effectively they require large-scale
adoption)~\cite{clark2013sok,matsumoto2015deployment}. For instance, even such a
critical functionality like certificate revocation cannot be successfully
deployed~\cite{liu2015end}.

We use a \textbf{blockchain as a publishing and contract enforcement
platform}.\footnotemark Although the blockchain technology is still in its
infancy, its inherent properties, like transparency, consistency, and censorship
resistance, make it a promising underlying technology for a notary system.
\footnotetext{Although our system does not necessarily require the blockchain
data structure and can be instantiated  with most distributed ledger systems, we
use the \textit{blockchain} term as the most of the existing distributed ledgers
deploy the blockchain data structure or similar.}

Firstly, \name notaries use the blockchain as a publishing platform to increase
transparency (auditability), availability, and to preserve users' privacy.  In
contrast to interactive client-server architectures, where clients have to
contact a notary to get its view, in \name notaries publish they validation
outcomes in the blockchain, thus anyone can read it \textit{passively}.

Secondly, smart contracts built upon the blockchain platform allow us to automate
the notary system by:
\begin{compactitem}
    \item introducing automated payments, that would incentivize notaries to
        keep a robust infrastructure, and
    \item implementing an SLA enforcement mechanism, to provide guarantees to the
        requesters (in the case of a misbehaving or unavailable notary).
\end{compactitem}
\medskip

We \textbf{decouple interface between notaries and requesters} into a
\textit{direct} and \textit{indirect} interface. Requesters and notaries follow
the client-server model for the direct interface,  which is used mainly for
auditing notaries (i.e., querying audit proofs).  However, for critical
operations, this model is enriched by the blockchain-based indirect interface,
where smart contracts act as a proxy to pass queries and as an oracle to resolve
potential disputes.  Operations over the indirect interface are implemented as
blockchain transactions thus are transparent, accountable, and cannot be
censored. This interface is intended only for critical operations like an
agreement creation or an SLA execution (e.g., when a notary is unresponsive or
censors queries sent via the direct interface), so interactions with
blockchain are involved only when it is necessary.

\subsection{High-level Overview}
\begin{figure}[bt]
    \includegraphics[width=\columnwidth]{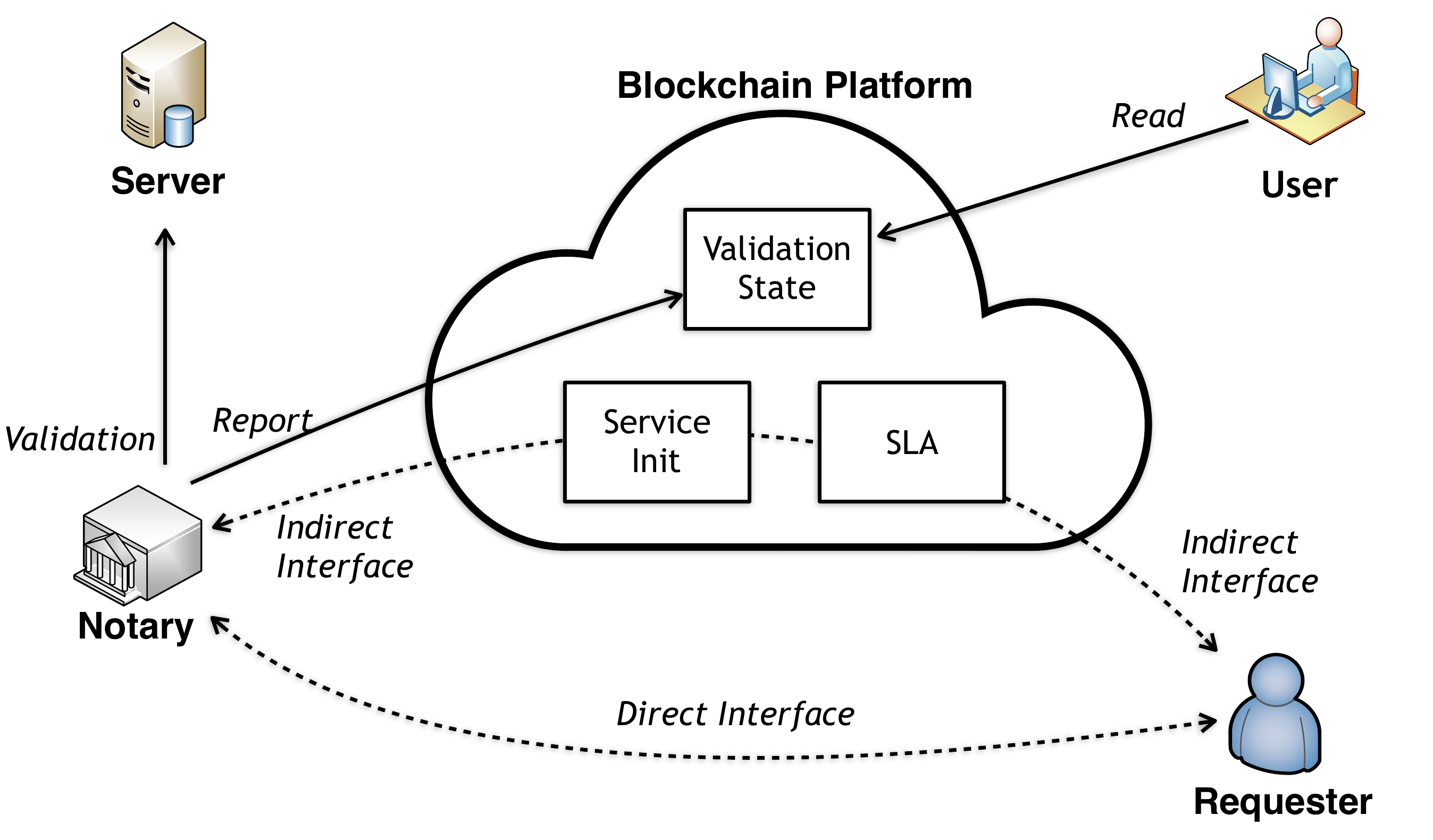}
    \caption{A high-level overview of \name.}
    \label{fig:overview}
\end{figure}

A high-level overview of \name is presented in \autoref{fig:overview}.  The
central point of our design is a blockchain platform that implements two main
functionalities (the indirect interface between notaries and requesters, and
publishing of validation results). The indirect interface is used to initialize
a \name service and to enforce an SLA between the notary and requester,
while the direct interface is used mainly for auditing the notary.

We assume that a requester knows (e.g., through a website) a notary, location
(address) of its smart contracts, the service fee, and the guaranteed SLA.  Then,
the protocol is executed as follows.
\begin{compactenum}
    \item In the first step, a requester requests the notary service by sending a
        transaction (to the notary's contract) that contains a fee for the
        service and specifies request details like a domain name to be validated
        and its legitimate public keys.
    \item The notary notices the request and configures the service to serve
        the request. Additionally, the notary deposits money that will be used
        to ensure its SLA.  Alternatively, the notary can refuse the request for
        any reason, like a misconfigured or inaccessible target server.
    \item Every \textit{validation period} $T$, the notary contacts the server
        and obtains an authenticated and fresh information about the server's
        public key.  Depending on an outcome of this validation the notary is
        obligated to publish in the blockchain any change of the \textit{validation
        state}.  Hence, the validation state can be read by any blockchain user.
    \item If there is no state change, the notary does not publish anything,
        which implies that the previous state is still valid.  Therefore, we
        optimize for the common case (i.e., no state change per $T$).
    \item In order to ensure that the validation state in the blockchain is
        correct,  at any time the requester can query the notary through the
        direct interface to provide validation result(s).  If the notary
        cannot provide the requested result(s) (e.g., the notary is unavailable
        or refusing the query), then the requester can use the indirect
        interface to send the same query over the blockchain. If within a
        predefined time period the notary does not respond using the indirect
        interface, then the SLA is executed (i.e., the deposit is sent to the
        requester and the service contract terminates).
\end{compactenum}
\medskip

The service in \name can be ordered by anyone, although we predict that usually
server operators have incentives to order it.  We do not mandate how validation
results are used and interpreted, but some potential deployment models can be:
\textit{a)} a public service accessible to all TLS clients, \textit{b)} a
monitoring tool used only by domain operators, \textit{c)} results can be
aggregated and delivered to TLS clients~\cite{larisch2017crlite}, or \textit{d)}
or can be part of a more powerful system~\cite{holz2012x}.

\section{\name Details}
\label{sec:details}
In this section, we present details of \name.  We start with service setup, then
we present how notaries validate public keys and report outcomes of these
validations. Next, we show how to monitor TLS servers that do not have clocks
synchronized, and finally, we present auditing and SLA mechanisms provided. 

\subsection{Service Setup}
\label{sec:detail:setup}
To order a notary service, a requester sends a transaction to the notary's contract
specifying the following parameters (see \texttt{request()} in
\autoref{fig:srv_setup}).
\begin{compactdesc}
    \item[Domain name] whose public key(s) will be validated.
    \item[Whitelist of public key(s)] that allows the requester to specify which
        public keys are legitimate.  If the notary will observe a listed key,
        then it is treated as a normal event and no warning notification is
        issued.  Keys in the list are identified by their hashes.  If the
        whitelist is empty, then any new public key observed will be reported.
    \item[Fee] that is paid to the notary for the service. Fees are predefined
        by notaries and determine the service duration and the SLA provided.
    \item[Reference time source] specifies a time source server that will be
        used as a reference.  This parameter is optional and used if the
        monitored server does not return the correct time.  If the time source
        is not provided, then the monitored server act as the reference time
        source.
\end{compactdesc}
\medskip

Beside parameters given by requesters, each service has associated conditions of
the SLA it offers.  These conditions specify:
\begin{compactitem}
    \item how frequent the notary validates the public key of the server (the
        validation interval $T$), and
    \item the availability level provided and a penalty when the level is not met.
\end{compactitem}
\medskip

The notary observes all transactions sent to its contract and processes incoming
requests (calling \texttt{handle\_request()} from \autoref{fig:srv_setup}). The
notary checks whether the requester's input is correct (checking the domain
name, the whitelist, the fee, and the time source if specified) and accepts the
request (\texttt{accept()} from \autoref{fig:srv_setup}) by creating a new
service and transferring deposit that will be used for ensuring the SLA.

The notary may refuse the request if the given parameters are incorrect, the
server is unresponsive or misconfigured, or for any other reason. In such a
case, the requester gets the fee back by calling the \texttt{timeout\_req()}
function that triggers the \texttt{timeout()} method (see
\autoref{fig:srv_setup}).

\begin{figure}[t!]
    \small
\begin{lstlisting}[language=python,frame=single]
# Requester initiates the service
def request(contract, domain, keys,
            fee, time_source=None)
  req = Request(domain, keys, fee,
                time_source)
  contract.request(req)

# The main contract class
class NotaryContract():
  # Schedule request
  def request(self, req):
    self.add_request(req)
    self.deposit(req.fee)

  # Accept request
  def accept(self, req):
    assert req in self.requests
    assert sender == self.owner
    self.deposit(self.SLA_DEPOSIT)
    self.add_service(req)
    self.remove_request(req)

  # Timeout pending request
  def timeout(self, req):
    assert req in self.requests
    assert sender == req.requester
    self.transfer(req.requester, req.fee)
    self.remove_request(req)
  ...

# Notary accepts the request
def handle_request(contract, req):
  # Validate the request parameters
  # like domain name, keys, etc ...
  assert validate(req)
  contract.accept(req)

# Requester triggers request timeout
def timeout_req(contract, req):
    contract.timeout(req)
\end{lstlisting}
    \caption{Pseudocode of the service setup operations.}
    \label{fig:srv_setup}
\end{figure}

\subsection{Public-Key Validation and Reporting}
\label{sec:details:validation}
\name is designed to be compatible with the current TLS infrastructure.
However, there are a few challenges associated with that.  Since we want to keep
notaries auditable and accountable, a notary to prove that it is validating a
domain's public key has to periodically obtain an authentic and fresh
information that the corresponding private key is being used by the domain.  But
the TLS protocol was not designed to provide such an information.

To overcome this issue, we exploit a message of the TLS handshake protocol (see
\autoref{sec:pre:tls:handshake}).  Namely, the \texttt{ServerKeyExchange}
message is signed with a server's private key, and is sent by a server when the
key exchange protocol negotiated is a variant of the \texttt{DHE\_DSS} or
\texttt{DHE\_RSA} methods. These methods are widely deployed and
used~\cite{adrian2015imperfect}. 

The \texttt{ServerKeyExchange} message is introduced to provide authentication
for client-server negotiated parameters.  It contains a server's signature over
the \texttt{signed\_params} structure, which consists of client and server
random values and server DH parameters (see \autoref{fig:params}).
\begin{figure}[h!]
\center
\begin{boxedverbatim}
digitally-signed struct {
        opaque client_random[32];
        opaque server_random[32];
        ServerDHParams params;
    } signed_params;
\end{boxedverbatim}
    \caption{The \texttt{signed\_params} structure, which is signed by a
    server during a TLS handshake.}
    \label{fig:params}
\end{figure}

This is the only message signed by a server during the TLS handshake, hence only
that can be obtained by notaries as a non-repudiable proof that a given key was
used.  However, the authenticity of the message is not enough, as it only proves
that the key was used, but does not specify when it happened.  

A message has to be fresh, as otherwise, a misbehaving notary could conduct
multiple handshakes at once, and then use them to claim a fake event timeline.
There is no explicit timestamp in the structure, but fortunately, the client and
server random inputs are defined as presented in \autoref{fig:rnd}, where the
\texttt{gmt\_unix\_time} field is the current GMT Unix 32-bit timestamp,
and the \texttt{random\_bytes} is 28-byte long random value, generated by a
secure random number generator.
\begin{figure}[t!]
\center
\begin{boxedverbatim}
struct {
        uint32 gmt_unix_time;
        opaque random_bytes[28];
    } Random;
\end{boxedverbatim}
    \caption{The \texttt{Random} structure, that specifies server's and client's
    random inputs.}
    \label{fig:rnd}
\end{figure}

Using these messages, the notary can periodically obtain a signed statement from
the server which implies that a given key was used at given point in time.  To do
so, the notary at least every $T$ conducts a new TLS handshake with the server
and saves as a \textit{validation result} the server's certificate, signed
values (including server's timestamp), and the signature.  These messages are
sent during a TLS handshake in plaintext, and they allow to prove to anyone that
the corresponding private key was used at the time specified in the
\texttt{gmt\_unix\_time} field of the \texttt{Random} structure sent by the
server.  For each such a public-key validation a notary assigns a unique number
(counter) called a \textit{validation id} (vid).  Validation results,
accompanied by their corresponding vids, are stored by notaries for audits.

A notary has to inform about results of the validation process.  A naive
approach would be to just to publish a validation result every $T$ in the
blockchain.  However, that would cause a significant overhead incurred by
publishing the result of every validation.  To minimize the interactions between
notaries and the blockchain (i.e., to not flood the blockchain with the protocol
messages), notaries keep all validation results but report in the blockchain
only changes of the validation state.  More specifically, the validation state
is encoded using the following messages:
\begin{compactdesc}
    \item[\texttt{OK}:]  successful validation, i.e., the observed public key
        is whitelisted.
    \item[\texttt{Error}:] unsuccessful validation that can be caused by various
        reasons. The following error types are specified:
    \begin{compactdesc}
        \item[\texttt{NewKey}:] a new public key, outside the whitelist, was
            observed. This event can denote a misconfiguration or an
            impersonation attack.  The notary that changes the state to
            \texttt{NewKey} publishes also the hash of the newly observed key.
        \item[\texttt{Time}:] timestamp signed by the monitored server (or the
            time source, if specified --- see the next section) in the
            \texttt{ServerKeyExchange} message is incorrect (i.e., deviating
            from the notary time).
        \item[\texttt{Connect}:] denotes availability issues, when the notary
            cannot conduct a successful TLS connection with the server.  It can
            be caused by a network outage, a server-side error, or an adversary
            blocking the connection.
        \item[\texttt{Other}:] other error, not specified above.
    \end{compactdesc}
\end{compactdesc}
\medskip

The notary submits each state update by calling \texttt{change\_state()} that
triggers the \texttt{state()} method of the notary's contract (see
\autoref{fig:state_change}).  A state update is associated with the
corresponding vid, i.e., the id of the validation when the state change happened.
If there is no state change, then no message is sent.  Using this simple
approach, \name optimizes for the common case (i.e., a public key of a server
changes infrequently).

\begin{figure}
    \small
\begin{lstlisting}[language=python,frame=single]
# The main contract class
class NotaryContract():
  ...
  # Change state
  def state(self, srv, new):
    assert srv in self.services
    assert sender == self.owner
    if (new.vid > srv.state.vid and
        new.status != srv.state.status):
      srv.state = new
  ...

# Notary changes the validation state
def change_state(contract, status, vid):
  new = State(status, vid)
  contract.state(srv, new)
\end{lstlisting}
\caption{Pseudocode of the state update procedure.}
    \label{fig:state_change}
\end{figure}
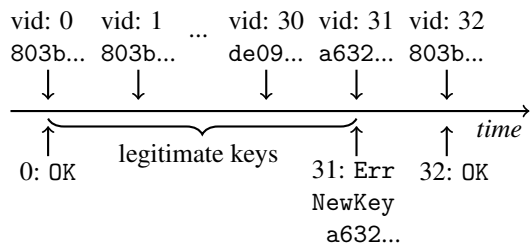
\begin{figure}[b!]
\centering
	\begin{tikzpicture}[scale=1]
    \scriptsize
	\draw [thick,->] (0,0) -- (6.9,0);
	\node[align=center] at (6.5,-0.225) {\textit{time}};

    \node[align=left] at (0.5,1) {vid: 0\\\texttt{803b}...};
	\draw [thick,->] (0.5,0.55) -- (0.5,0.15);

    \node[align=left] at (1.7,1) {vid: 1\\\texttt{803b}...};
	\draw [thick,->] (1.7,0.55) -- (1.7,0.15);

    \node[align=left] at (2.5,1) {...};

    \node[align=left] at (3.4,1) {vid: 30\\\texttt{de09}...};
	\draw [thick,->] (3.4,0.55) -- (3.4,0.15);

    \node[align=left] at (4.6,1) {vid: 31\\\texttt{a632}...};
	\draw [thick,->] (4.6,0.55) -- (4.6,0.15);

    \node[align=left] at (5.8,1) {vid: 32\\\texttt{803b}...};
	\draw [thick,->] (5.8,0.55) -- (5.8,0.15);
	\draw [thick,decorate,decoration={brace,amplitude=6pt,raise=0pt,mirror}]
        (0.5,-0.1) -- (4.6,-0.1);
	\node[align=center] at (2.5,-0.6) {legitimate keys};

	\draw [thick,->] (0.5,-0.6) -- (0.5,-0.20);
    \node[align=left] at (0.5,-0.8) {0: \texttt{OK}};

	\draw [thick,->] (4.6,-0.6) -- (4.6,-0.20);
        \node[align=left] at (4.6,-1.2) {31: \texttt{Err}\\\texttt{NewKey}\\\texttt{ a632}...};

	\draw [thick,->] (5.8,-0.6) -- (5.8,-0.20);
    \node[align=left] at (5.9,-0.8) {32: \texttt{OK}};

	\end{tikzpicture}
    \caption{Timeline of the validation results and the corresponding state
    changes published in the blockchain. In the example, the public-key
    whitelist is set to \{\texttt{803b}..., \texttt{de09}...\}.}
    \label{fig:time}
\end{figure}

Reported state changes allow drawing an event timeline. An example of such a
timeline is presented in \autoref{fig:time}.  In the first validation, the
notary obtains a whitelisted public key, thus it reports the \texttt{OK} message
with vid equals 0.  Then, the subsequent 30 validations are also successful,
thus no state change is published in the blockchain. With the validation 31, the
notary detects a public key outside the whitelist, therefore a \texttt{NewKey}
error specifying the hash of the newly observed key is reported.  The last
validation presented is again successful, hence the notary changes the state
back to \texttt{OK}.

\subsection{Handshake Timestamping}
\label{sec:details:time}
So far, we present \name in the setting where a timestamp returned by the
monitored server acts as the reference time.  However, the TLS protocol does not
require clocks to be set correctly~\cite{rfc5246}, and moreover, some
implementations violate the specification and do not set correct
timestamps~\cite{remove_gmt}.  As presented in \autoref{sec:impl:eval}, about
63\% of tested TLS servers do not put correct timestamps into their
\texttt{ServerHello} messages.

That could limit deployment of \name since TLS servers that do not return
correct timestamps cannot be reliably monitored (without server-side timestamp
it would be impossible to tell when a given validation happened).  To overcome
this issue, \name allows notaries and requesters to agree on a reference time
source (see \autoref{sec:detail:setup}) which will be used for timestamping TLS
handshakes.  For simple description, we assume that a time source server for any
client's input returns this input timestamped and signed.
\begin{figure}[b!]
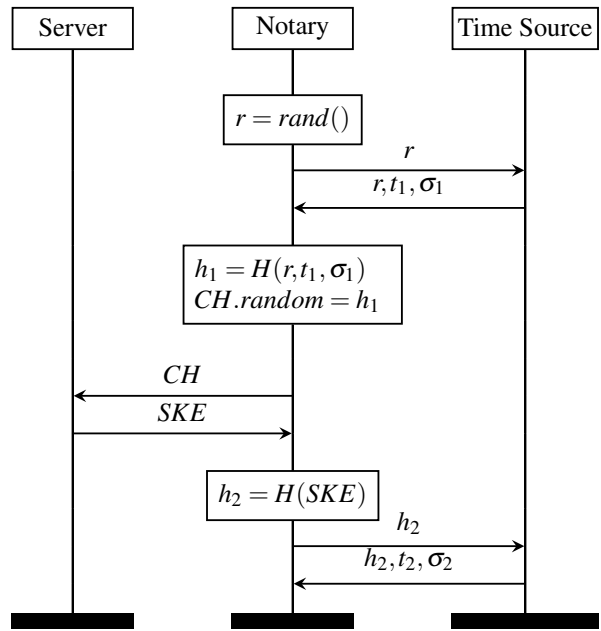

    \vspace{-1cm}
    \centering
    \drawframe{no}
    \setmsckeyword{}
    \setlength{\envinstdist}{0.5cm}
    \setlength{\instdist}{1.3cm}
    \begin{msc}{}
        \small
        \declinst{ser}{}{Server}
        \declinst{not}{}{Notary}
        \declinst{ts}{}{Time Source}

        \action*{$r=rand()$}{not}
        \nextlevel
        \nextlevel
        \mess{$r$}{not}{ts}
        \nextlevel
        \mess{$r,t_1,\sigma_1$}{ts}{not}
        \nextlevel
        \action*{\parbox{2.6cm}{
            $h_1=H(r,t_1,\sigma_1)$\\
            $CH.random=h_1$
        }}{not}
        \nextlevel
        \nextlevel
        \nextlevel
        \nextlevel
        \mess{$CH$}{not}{ser}
        \nextlevel
        \mess{$SKE$}{ser}{not}
        \nextlevel
        \action*{$h_2=H(SKE)$}{not}
        \nextlevel
        \nextlevel
        \mess{$h_2$}{not}{ts}
        \nextlevel
        \mess{$h_2,t_2,\sigma_2$}{ts}{not}
    \end{msc}
    \vspace{-0.5cm}
    \caption{The extended public-key validation protocol, with the time source
    server involved. $H()$ stands for a cryptographic hash function.}
    \label{fig:time_valid}
\end{figure}

To prove that the public-key validation happened at the given time (according to
an external time source) we introduce a protocol that the notary executes at
every validation.  The protocol is presented in \autoref{fig:time_valid} and its
steps are executed as follows:
\begin{compactenum}
    \item the notary initiates the protocol by sending a random value $r$ to the
        time source server.
    \item The time source, responds with a signed message $(r,t_1,\sigma_1)$
        which contains the random input, a timestamp $t_1$ (set by the time
        source), and the signature computed by the time source.
    \item Next, the notary prepares the \texttt{ClientHello} message $CH$ to be
        sent to the monitored server. The notary sets the \texttt{random} field
        of this message to the hash of $(r,t_1,\sigma_1)$, and subsequently
        initiates a TLS handshake with the monitored server by sending $CH$.
    \item The server responds with the signed \texttt{ServerKeyExchange} message
        $SKE$.  Timestamp included in this message is incorrect (as we assumed
        that the server timestamps are unreliable), however, the
        \texttt{client\_random} field is the hash over the $(r,t_1,\sigma_1)$
        message, which in turn contains the trusted timestamp $t_1$. Thus, it
        implies that $SKE$ was created after $t_1$.
    \item Then, the notary sends $h_2=H(SKE)$ to be timestamped by the time
        server.
    \item Finally, the time source returns a signed $(h_2,t_2,\sigma_2)$
        message, which contains the (trusted) timestamp $t_2$ and $H(SKE)$ in
        the \texttt{client\_random} field. This implies, that $SKE$ is older
        than $(h_2,t_2,\sigma_2)$, thus is older than its timestamp $t_2$.
\end{compactenum}
\medskip

By executing the protocol, the notary gets the evidence that the message $SKE$
signed by the monitored server is older than $(r,t_1,\sigma_1)$ and newer than
$(h_2,t_2,\sigma_2)$.  The notary keeps all the messages of the protocol as a
validation result, that can prove that the server's public key was validated
between $t_1$ and $t_2$. 

Interestingly, any TLS server that can be monitored by \name, can also act as a
time source.  In such a case, the notary simply uses the random field of the
\texttt{ClientHello} message to submit data to be timestamped and receives a
\texttt{ServerKeyExchange} message that signs this data.  As presented in
\autoref{sec:impl:eval}, in such a setting an average time difference between
receiving $t_2$ and $t_1$ was measured as 0.34s, thus such estimation is
precise.  Moreover, the protocol can be also easily combined \name with other
protocols and infrastructures (as discussed in \autoref{sec:disc:time}).

\subsection{Auditing and SLAs}
The main task of a notary is to periodically check a monitored server's public
key(s). \name keeps notaries auditable and accountable for their actions, and to
facilitate the audit process, a notary is obligated to:
\begin{compactitem}
    \item publish all changes of the validation state in the blockchain (as
        described in \autoref{sec:details:validation}),
    \item store validation results for all monitored domains throughout their
        service lifetimes.
\end{compactitem}
\medskip

At any time the requester can ask a notary about any validation
results.\footnote{In fact, all validation results can be publicly published by
notaries, such that everyone can audit it.}  This is realized with the direct
interface, implemented as a standard client-server communication (e.g., an HTTP
API), where the requester sends a query specifying the validation id for which
he would like to obtain the result.  The notary has to return the relevant
validation result, such that the requester can compare it with the state encoded
in the blockchain.

In the case of an inconsistency detected, the requester can publicly announce
the validation id that contradicts the state encoded in the blockchain, as it is
evidence that the notary misbehaved.  For instance, it can be a validation
result where the validated key is not whitelisted, while at the same time the
state in the blockchain was not updated (e.g., when any validation between vid 1
and 30 was for a non-whitelisted key --- see \autoref{fig:time}).

One challenge with the direct interface (and any client-server service) is that
the notary can be unavailable or censor queries while the requester cannot
prevent or prove that fact.  In our context, a notary can have some reasons to not
return some validation results. For instance, the notary is unavailable, or
was unavailable at some validation period(s), misbehaved, or made a mistake and
tries to hide that.  \name provides a framework to keep notaries accountable and
responsible for such cases.  Namely, besides the standard client-server direct
interface, the blockchain-based indirect interface is introduced. 

If the requester notices that the notary is unavailable (or was unavailable at a
given time interval) or censors queries, the requester can query the notary over
the blockchain.  In order to do so, the requester calls the \texttt{query()}
function that triggers the \texttt{sla\_query} of the notary's contract (see
\autoref{fig:sla}).  When the transaction with the query is added to the
blockchain it is managed by a smart contract and the notary is obligated to
submit a response to the blockchain.
\begin{figure}[h!]
    \small
\begin{lstlisting}[language=python,frame=single]
# Requester's query
def query(contract, srv, vid):
    q = Query(srv, vid)
    contract.sla_query(q)

# The main contract class
class NotaryContract():
  ...
  # Publish requester's query
  def sla_query(self, q):
    assert q.srv in self.services
    assert sender == q.srv.requester
    q.time = time()
    self.add_query(q)

  # Publish notary's response
  def sla_response(self, q, resp):
    assert q in self.queries
    assert q.srv in self.services
    assert sender == self.owner:
    self.add_response(resp)
    if time() - q.time <= self.SLA_TOUT:
      self.remove_query(q)

  # Execute SLA
  def sla_claim(self, q):
    assert q in self.queries
    assert q.srv in self.services
    assert sender == q.srv.requester
    if time() - q.time > self.SLA_TOUT:
      self.transfer(q.srv.requester,
                    self.SLA_DEPOSIT)
      self.remove_query(q)
      self.remove_service(q.srv)

# Notary's response
def response(contract, q, msg):
    resp = Response(q, msg)
    contract.sla_response(srv, q, resp)
\end{lstlisting}
\caption{Pseudocode of the SLA operations.}
    \label{fig:sla}
\end{figure}

If the notary does not return any result until the deadline specified in the SLA
then the smart contract will execute the SLA,
sending the deposit to the requester.  To trigger it, the requester
calls the contract's \texttt{sla\_claim()} method with the unresponded query
specified.  It is also visible to anyone that the notary did not respond to the
query. Alternatively, when the notary returns a response (calling the
\texttt{response()} function that triggers the contract's
\texttt{sla\_response()}), the SLA is not executed, as the notary
\textit{proved} its availability, and the response can be processed by the
requester (see details in \autoref{sec:analysis}).  Also, in that case, the
response is visible to everyone.

Thanks to dividing interactions into the two interfaces (the standard
client-server and the blockchain-based one) the regular audit operations are
conducted efficiently without involving blockchain, while the blockchain
interface is used only when necessary.

\section{Security Analysis}
\label{sec:analysis}
\paragraph{MitM attacks}
The main goal of notary systems is to detect and prevent MitM attacks
through multipath probing.  \name follows this strategy, except the public key
validation is positioned as a service, where notaries constantly validate public
keys and draw event timelines in the blockchain. These timelines are essential
for measuring key continuity and detecting anomalies. 

We do not mandate \name to be deployed in a specific way. Service can be
requested by domain owners and used only by themselves. Alternatively, it can be
run as a public service accessible for everyone or can be used as a data feed
for other systems~\cite{holz2012x,larisch2017crlite}.  We also emphasize that
\name is orthogonal to the TLS PKI. A requester can whitelist any public
keys believed or known to be legitimate.

In all these cases, \name can provide an effective protection against MiTM
attacks as long as an attack is short-lived or is limited only to a fragment of
the network topology.  In the former case, the attack is seen as an anomaly
(i.e., a new observed public key outside the whitelisted set), thus it is
suspicious.  In the latter case, the attack is identified as an inconsistency
between the notary and the TLS client views.  In this context, the previous
systems provide similar properties, however, the advantage of \name is that it
provides more reliable and auditable event timelines.

\paragraph{Misbehaving Notary}
\name keeps notaries accountable, and \textit{if an adversary impersonates the
domain throughout the validation interval, then the notary cannot make a false
statement undetected that it has not happened}.  More specifically, if the notary
contacts the monitored server and notices a public key that is not whitelisted
by the requester, then either \textit{a)} the notary announces the state change
making it visible, or \textit{b)} it does not report the state change, hiding
this fact.

In the latter case, the notary takes a risk, as the requester conducting an
audit will notice that either the state is incorrect (as one or more validation
results are contradicting it), or the validation result(s) are missing.  (We
emphasize that the notary cannot produce a validation result on behalf of the
monitored server, as the private key corresponding to a whitelisted key is
required.) If some validation results are contradicting the published state, the
requester has evidence of the notary misbehaving.  However, if validation
results are missing/censored, the requester can query the notary via the
indirect interface.  The notary is obligated to response, so either it responds
with an incorrect data, or it refuses to respond, losing the SLA deposit and
showing its unavailability to everyone.

\paragraph{(D)DoS Resilience}
Availability is one of the major issues of the previous notary systems. They
were designed in the client-server architecture, where a client wishing to learn
a validation state has to contact a notary server and wait for its response.
Client queries are expected to be handled immediately 24/7, and due to that
availability of notary servers is crucial.  Unfortunately, such a setup can be
easily attacked with a botnet targeting notary front-end servers, practically
stopping their operations.

\name provides a more flexible architecture, by placing a blockchain as a
publishing medium, while still exposing a client-server interface for
heavyweight operations.  An adversary with a botnet can still attack notary
front-end servers blocking their direct interfaces, however, it is much more
difficult to stop the notary's operation. The notary can still use any back-end
server (unknown to the adversary) to conduct public-key validations,  and report
the corresponding state changes (if any) in the blockchain, such that it is
available to everyone who reads the blockchain.  Namely, \name changes
availability requirements from \textit{``a notary needs to be up all the time''}
to \textit{\textbf{``a notary has to have at least one back-end up once per the
validation period $T$''}}.  Moreover, to avoid enumeration back-ends can be in
different locations, with dynamic IPs, using network tunnels or anonymity
infrastructures.  To completely block notary operations, the adversary has to
either block the underlying peer-to-peer blockchain network or censor blockchain
transactions.

\paragraph{Privacy}
\name provides privacy benefits over the previous schemes, as the validation
state history can be read from the blockchain, without contacting any third
party.  In particular, no notary infrastructure is contacted by clients.

\section{Implementation and Evaluation}
\label{sec:implementation}
To prove deployability and efficiency of \name we implemented it and then
evaluated in a real-world scenario.  In this section, we report our
implementation and the obtained evaluation results.

\subsection{Implementation}
\begin{figure}[h!]
    \centering
    \includegraphics[width=0.86\columnwidth]{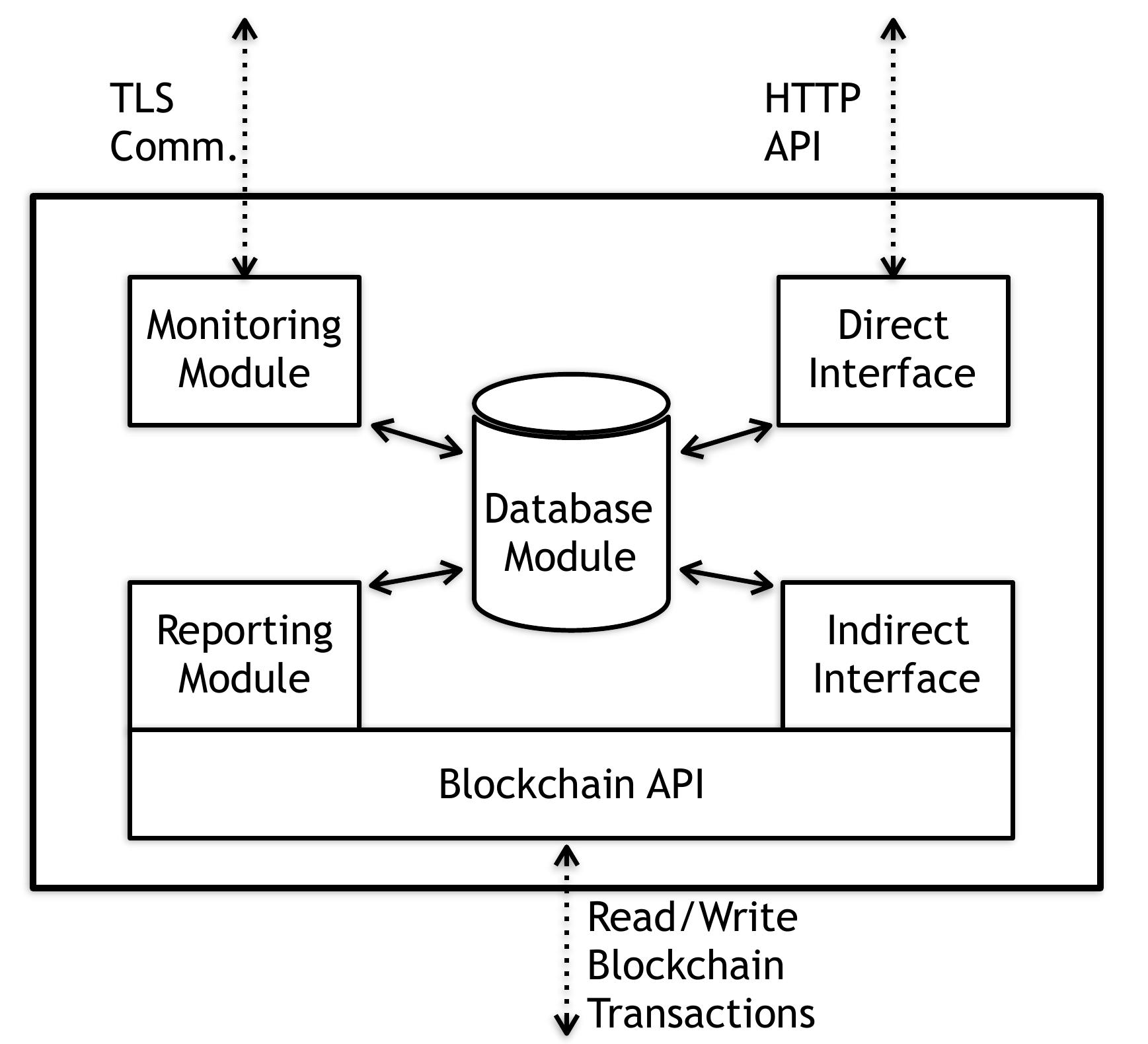}
    \caption{The internal architecture of a \name notary.}
    \label{fig:notary_arch}
\end{figure}

The notary's internal architecture is presented in \autoref{fig:notary_arch} and
it consists of the following elements:
\begin{compactdesc}
    \item[Monitoring module] periodically conducts TLS handshakes with monitored
        domains. This module collects cryptographic proofs and supports the
        timestamping protocol (as described in \autoref{sec:details:time}).
        Results gathered by the monitoring module are reported to the database
        module.
    \item[Database module] stores all necessary state and data to run a notary
        service (i.e., active and inactive services, results of all validations,
        pending requests, ...).
    \item[Reporting module] observes state of validations and reports
        validation state changes into the blockchain (using the blockchain API).
    \item[Direct interface] is an HTTP API used to serve validation results to
        requesters. The direct interface can be publicly accessible to provide
        this data to any party that would like to audit the notary.
    \item[Indirect interface] uses the blockchain API to handle service
        initialization and SLA requests.
\end{compactdesc}

We implemented most of the modules in Python.  The monitoring module was
implemented with the Scapy and Scapy-SSL-TLS libraries. In our implementation
TLS handshakes are not finished, instead, a notary sends a \texttt{TCP RST}
packet whenever a \texttt{ServerHelloDone} message is received (see
\autoref{fig:tls_handshake}).  With this modification, our implementation
saves one round-trip time and is server-friendly (as after receiving a
\texttt{TCP RST} the server removes the associated connection state).

We chose Ethereum as the blockchain platform, and the smart contract part of
\name was implemented using the Viper
language\footnote{\url{https://github.com/ethereum/viper}}.  The blockchain API
was implemented with Go's\footnote{\url{https://geth.ethereum.org/}} and
Python's\footnote{\url{https://github.com/ethereum/pyethereum}} Ethereum
implementations.

\subsection{Evaluation}
\label{sec:impl:eval}
First, we investigated what is the fraction of TLS servers can be monitored by
\name.  To this end, we scanned 15\,000 domains from the Alexa top
list\footnote{\url{http://s3.amazonaws.com/alexa-static/top-1m.csv.zip}} and
checked how many of them deploy TLS on the HTTPS port (i.e., 443). If a domain
does not deploy TLS we also try to prepend it with the ``\texttt{www.}'' prefix.
Then, for the deploying domains, we checked how many of them return correct
timestamps in the \texttt{ClientHello} message (prior to this test we
synchronized our clock with \url{time.google.com}).  The results are presented
in \autoref{tab:tls-time}.  As presented, about 37\% of the TLS servers
investigated return precise timestamps (i.e., deviating from the correct time up
to one second).  These servers can be monitored by \name notaries directly
(without involving a time source) or can act as time sources for servers that
return incorrect timestamps.

\begin{table}[t!]
\begin{center}
    \small
    \begin{tabular}{l|rrrrr}
        $\Delta$ (s) & 0-1    & 2-5    & 6-60   & 61-300 & $>$300 \\ \hline
        \# of        & 3\,061   & 420    & 130    & 82     & 5\,823 \\
        servers      & 32.17\%& 4.41\% & 1.37\% & 0.86\% & 61.19\%\\
\end{tabular}
\end{center}
    \caption{The distribution of timestamps returned by the TLS servers. $\Delta$ is
    defined as $abs(t_l-t_s)$, where $t_l$ is the current local time, and
    $t_s$ is a server's timestamp.}
    \label{tab:tls-time}
\end{table}

Next, we conducted a series of experiments to evaluate our implementation of
\name in a realistic scenario. We deployed a notary server using Linux Ubuntu
16.04 (64 bit) equipped with Intel i7-6600U CPU @ 2.60GHz and 8GB of RAM.  From
an academic network in Asia, we conducted 100 public-key validations selecting a
random supported TLS server (the \textit{single-server} case) each time, and
another 100 public-key validations where for each validation we selected a
random supported TLS server and another supported TLS server acting as a time
source (the \textit{timestamp} case).  We reported the obtained results in
\autoref{tab:overheads}, specifying minimum, maximum, average, and median value
for every measurement.

\begin{table}[b]
\begin{center}
    \small
    \begin{tabular}{l|rrrr}
                       &   Min  &   Max   &   Avg.  &  Med.   \\ \hline
        $t_{sing}$ (s) &   0.04 &    0.31 &    0.16 &    0.15 \\
        $t_{ts}$ (s)   &   0.19 &    0.73 &    0.50 &    0.50 \\
        $s_{sing}$ (B) & 1\,313 &  6\,865 &  4\,369 &  4\,483 \\
        $s_{ts}$ (B)   & 6\,031 & 17\,973 & 13\,190 & 13\,445 \\
        $t_2-t_1$ (s)  &   0.16 &    0.48 &    0.34 &    0.34 \\
        $s_{cert}$ (B) &    806 &  6\,071 &  3\,815 &  3\,976 \\
    \end{tabular}
    \end{center}
    \caption{The obtained performance results.}
    \label{tab:overheads}
\end{table}

For both, the single-server and timestamp case, we report time required to
conduct a validation ($t_{sing}$ and $t_{ts}$ respectively), as well as
transmission overhead incurred in the validation ($s_{sing}$ and $s_{ts}$
respectively).  The validation time is mainly determined by network latency.
(We emphasize, that in our implementation only \textit{half} TLS handshakes are
conducted, minimizing this latency.) For the single-server case, it ranges
between 0.04-0.31s with an average value of 0.16s. For the same case, the number
of bytes transmitted is between 1.3-6.9KB, with an average around 4.5KB.  The
timestamp case, which requires three TLS handshakes, needs between 0.19s to
0.73s with an average value of half a second, and to transmit between 6-17KB
with an average of about 13.5KB.

Next, we investigated how precise is the variant of the public-key validation
with time source server involved. To this end, we measured the time difference
between receiving the timestamps $t_1$ and $t_2$ from a time server (see
\autoref{fig:time_valid}).  This measurement illustrates how much time it takes
to timestamp a TLS handshake (i.e., what overhead is introduced by running the
protocol from \autoref{fig:time_valid}).  The results obtained by our
measurements are presented in the table.  As we can see, the overhead introduced
by deploying the timestamping protocol is around 0.20s on average, what should
be marginal when compared with a realistic validation period $T$ (see
\autoref{sec:block_time}).

\name notaries have to store exchanged TLS messages as validation results.
However, when stored naively this storage overhead may be significant.  On
average, a single validation takes 4\,483B, thus a storage required for yearly
validations of one server, with $T$ equals one hour, would be around 39.27MB.
However, validation results are usually highly redundant, as their size is
dominated by certificates sent from servers to notaries (see
\autoref{fig:tls_handshake}).  Since certificates are usually the same, there
is no need to store them all.  As an average certificate chain's size $s_{cert}$
was measured as 3\,976B, the same yearly validation results without storing
redundant certificates would require 4.44MB (almost 9 times less).

\section{Discussion}
\label{sec:discussion}
\subsection{Time Sources}
\label{sec:disc:time}
We presented \name such that it deploys TLS servers as external time sources
(see \autoref{sec:details:time}). However, our scheme can be easily combined
with other protocols and infrastructure that provide authentic and reliable
timestamps.

One, such an infrastructure is the time-stamp protocol (TSP)~\cite{rfc3161}, a
document timestamping protocol that relies on the X.509 PKI.  In TSP, a client
submits a hash of data to a timestamp authority which in turn timestamps and
signs this data. The signed message is returned to the client, so the client can
prove to everyone when the data was timestamped by the authority.  This model is
almost the same as the model presented in \autoref{sec:details:time} (see
\autoref{fig:time_valid}).  Thus, a TSP server can be used as an external time
source in \name.

Another example is time synchronization infrastructures. Usually, secure time
synchronization protocols use random client inputs to prevent replay attacks.
Such an input is timestamped and signed by a time synchronization server, thus
it can be used analogically as presented in \autoref{fig:time_valid}.  One
instantiation of this approach could be a novel time synchronization protocol
Roughtime proposed by
Google.\footnote{\url{https://roughtime.googlesource.com/roughtime}}

\subsection{TLS 1.3}
As for February 2018, the next version (1.3) of the TLS protocol is being
standardized.  The standardization process is not finished and as we have
learned from other server-side protocol upgrades~\cite{nykvistserver}, we should
not expect a quick upgrade to TLS 1.3 (especially, as the TLS 1.3 failure rate
is high right now~\cite{haztls13,whytls13}).  However, with the current
draft~\cite{rescorla2016transport} the TLS 1.3 protocol introduces the following
changes which might affect \name:
\begin{compactenum}
    \item removed the \texttt{ServerKeyExchange} message type,
    \item removed the GMT timestamp fields from the client and server random
        structures.
\end{compactenum}

The first change is not disruptive, as the new \texttt{CertificateVerify}
message with similar semantics was introduced instead. The
\texttt{CertificateVerify} message contains a signature that is computed over
the client's and server's random inputs, thus it can be used to prove that a
given key was used for signing (exactly as with the \texttt{ServerKeyExchange}
message).

The latter change was introduced to prevent fingerprinting, as a GMT timestamp
could be used to track clients and servers (note, that the most of the handshake
messages are exchanged in the plaintext, thus an eavesdropping adversary can
read them)~\cite{remove_gmt}. Due to this change, it is impossible to prove when
a given key was used (as there is no timestamp). If this change reminds stable
in the specification, it will break relying upon protocols (like
TLSdate\footnote{\url{https://github.com/ioerror/tlsdate}}) and will partially
affect \name as well.  Namely, the simple version (the version without external
timestamping presented in \autoref{sec:details:validation}) of the protocol will
become unusable as due to this change notaries would not be able to prove when a
given key was used.  Fortunately, \name can still be deployed in the scenario
with an external time source (see \autoref{sec:details:time}). Moreover, if the
monitored server deploys TLS 1.3, then any server with a lower TLS version
(e.g., TLS 1.2) can act as a time source.  Furthermore, as described in
\autoref{sec:disc:time} other services can be used for providing reference
timestamps.

\subsection{Blockchain Time}
\label{sec:block_time}
A public blockchain is an open distributed infrastructure, and nodes do not have
precisely synchronized clocks~\cite{szalachowski2018towards}.  However, by
definition, the blockchain structure preserves the order of blocks, and this is
why time in the presented smart contracts (i.e., \texttt{time()} calls) can be
expressed in numbers of blocks and not in seconds (usually, blockchain platforms
are configured such that new blocks arrive in equal intervals).  Consequently,
timeouts specified in SLAs can be measured in block numbers.

The validation interval $T$ is specified by a notary in seconds.  It is a
trade-off between security and efficiency, and although we do not require any
minimum/maximum values for $T$, there are some underlying constraints and
implications that notaries should take into consideration.  In particular, \name
has to combine the two notions of time: \textit{a)} time maintained by TLS
servers or external time sources, used for conducting validations, and
\textit{b)} time of the underlying blockchain platform.

In blockchain systems new blocks arrive at intervals (e.g., approximately 10
minutes in Bitcoin~\cite{nakamoto2008bitcoin}, and 15 seconds in Ethereum).  Let
us denote this interval as $T_B$.  If $T<T_B$, then reporting multiple state
changes within $T_B$ has some implications as it can occur that many state
changes happen within $T_B$, such the corresponding transactions announcing it
will be added in the same block.  Each state change has the associated
validation id, thus as the results, the timeline can be created, however, users
that observe the blockchain get results in batches.  With $T\geq T_B$ this
problem is relaxed as usually state changes will be propagated in separate
blocks.

\section{Related work}
\label{sec:related}
Perspectives~\cite{wendlandt2008perspectives} was the first comprehensive notary
system presented.  In Perspectives, notaries continuously observe domain
certificates. Clients can contact a notary server and compare their view of the
domain's key with the view of the notary.
Convergence~\cite{marlinspikeconvergence} is a similar system where the main
improvements over Perspectives are related to privacy and performance.
Deployment of Convergence was analyzed by Bates et al.~\cite{bates2014forced},
where the increased latency of TLS connection establishment (about 108 ms) is
reported.  Besides that, these systems are critiqued as they need to have highly
available front ends, they require a significant trust in notaries, and
introduce privacy violations~\cite{not_convergence,merzdovnik2016whom}.

Certificate Transparency (CT)~\cite{rfc6962} is a log-based detection scheme.
CT aims to introduce transparency to the CA ecosystem, by making each
certificate visible. To this end, CT introduces publicly verifiable certificate
logs that maintain append-only databases of certificates.  To accept a TLS
connection, the client has to obtain a certificate accompanied by a signed
promise from a log (the promise asserts that certificate was seen by the log).
CT is designed to detect attacks, rather than prevent them. Logs are required
for certificate issuance, thus their front-end servers have to be highly
available, what in practice is
challenging~\cite{ct_uptime,ct_uptime2,ct_old_sth}.  Although log servers
are auditable, to assure that they do not equivocate (by creating an
alternative log's version) they need to be monitored by external
protocols~\cite{chuat2015efficient,nordberg2015gossiping}.  \name aims to
improve CT ensuring that a given key is actually being used,
moreover, the availability requirement is significantly relaxed, and notaries
cannot equivocate as long as the underlying blockchain platform is secure.

CT has inspired other log-based approaches, that improve its
efficiency~\cite{ryan2014enhanced,kim2013aki}, enhance
security~\cite{kim2013aki,basin2014aar}, and add new
features~\cite{ryan2014enhanced,PKISN}. Unfortunately, some of these system
increase latency of TLS connection establishment, and other require major
changes to the TLS PKI.  As for today, no other log-based system than CT is
widely deployed.

CoSi~\cite{syta2016keeping} is a witness framework proposed to keep CA
accountable.  In CoSi, a large number of witnesses co-signs assertions about
certificates they have seen.  As a witness scheme, CoSi is focused on detection
rather than prevention, and its design objective (similarly to CT) is
to make attacks visible. It is assumed that an adversary would not conduct an
attack if it would be eventually visible.  In order to achieve efficient
co-signatures, CoSi requires coordination among witnesses.

Researchers already tried to reuse properties of blockchain to problems existing
in PKI.  For instance, Namecoin~\cite{loibl2014namecoin} (the first fork of
Bitcoin) provides a namespace where names can be
associated with public keys.  Certcoin~\cite{fromknecht2014certcoin} improves
Namecoin's inefficiency and supplements the original scheme by features like key
revocation and recovery.  Bonneau proposed EthIKS~\cite{bonneau2016ethiks} to
improve auditability of the CONIKS system~\cite{melara2015coniks}, while
Matsumoto and Reischuk~\cite{Reischuk16:IKP} introduce a blockchain-based system
that provides financial incentives for detecting fraudulent certificates.

\section{Conclusions}
\label{sec:conclusions}
In this paper, we presented \name, the next generation TLS notary system.  \name
provides many features and properties that the previous notary system could not
provide. By leveraging the TLS specification and redesigning the validation
process, notaries in \name do not have to be trusted as much as they were in the
past.  Notaries become auditable and accountable, able to monitor desynchronized
servers, and ready for TLS 1.3.  By placing a blockchain platform as one of its
central elements, \name provides a flexible payment framework and mechanisms for
defining and enforcing service-level agreements, and relaxes availability
requirements making the overall system more resilient.  The public-key
validation is persistent and implemented as a service.

With the increasing capabilities of blockchain platforms, \name could be
improved by more sophisticated smart contracts. For instance, with the current
design, a notary can respond via blockchain with any (even incorrect) response.
Although doing so the notary would ruin its reputation (as everyone can see
it), in the future the SLA contract could validate the response and act
accordingly.


\bibliographystyle{plain}
\bibliography{ref,rfc}
\end{document}